\documentclass[11pt]{article}

\usepackage{graphicx}
\usepackage{amssymb}
\usepackage{amsmath}
\usepackage{subfig}
\usepackage{hyperref}
\usepackage{tablefootnote}

\usepackage{natbib}

\usepackage{JASA_manu}

\begin{document}

\title{Fast Hamiltonian Monte Carlo Using GPU Computing}
\author{Andrew L. Beam$^1$, Sujit K. Ghosh$^{2,3}$, Jon Doyle$^{1,4}$  \\
$^1$Bioinformatics Research Center. North Carolina State University\\
$^2$Department of Statistics. North Carolina State University.\\
$^3$Division of Mathematical Sciences. National Science Foundation.\\
$^4$Department of Computer Science. North Carolina State University.\\
email: \texttt{albeam@ncsu.edu} }

\maketitle

\newpage

\mbox{}
\vspace*{2in}
\begin{center}
\textbf{Author's Footnote:}
\end{center}
Andrew L. Beam is a PhD candidate in Bioinformatics at North Carolina State University. \\
Dr. Ghosh is is a Professor in the Department of Statistics at North Carolina State University and currently serves as a Program Director in the Division of Mathematical Sciences (DMS) at National Science Foundation (NSF). \\
Dr. Doyle is the SAS Institute Distinguished Professor of Computer Science at North Carolina State University.\\


\newpage
\begin{center}
\textbf{Abstract}
\end{center}
In recent years, the Hamiltonian Monte Carlo (HMC) algorithm has been found to work more efficiently compared to other popular Markov Chain Monte Carlo (MCMC) methods (such as random walk Metropolis-Hastings)  in generating samples from a posterior distribution. A general framework for HMC based on the use of graphical processing units (GPUs) is shown to greatly reduce the computing time needed for Bayesian inference. The most expensive computational tasks in HMC are the evaluation of the posterior kernel and computing its gradient with respect to the parameters of interest. One of primary goals of this article to show that by expressing each of these tasks in terms of simple matrix or element-wise operations and maintaining persistent objects in GPU memory, the computational time can be drastically reduced. By using GPU objects to perform the entire HMC simulation, most of the latency penalties associated with transferring data from main to GPU memory can be avoided. Thus, the proposed computational framework is conceptually very simple, but also is general enough to be applied to most problems that use HMC sampling. For clarity of exposition, the effectiveness of the proposed approach is demonstrated in the high-dimensional setting on a standard statistical model - multinomial regression. Using GPUs, analyses of data sets that were previously intractable for fully Bayesian approaches due to the prohibitively high computational cost are now feasible.

\vspace*{.3in}


\newpage

\section{Introduction}
Bayesian methods continue to grow in popularity for a variety of data analyses and have many appealing properties, especially when the primary task is prediction. Bayesian methods offer a coherent and principled framework for model regularization that is often crucial in high-dimensional settings. Many popular frequentist approaches, such as the Lasso (\cite{tibshirani1996regression}) and Ridge Regression (\cite{hoerl1970ridge}), have been used extensively in high-dimensional settings are well known as approximating a Bayesian posterior distribution under various priors. Probabilistic inference based on the full posterior distribution using analytical or deterministic numerical methods in large scale problems is almost always intractable, so Markov Chain Monte Carlo (MCMC) techniques are often used to draw samples from the posterior distribution based on only the kernel of the posterior density. These approaches do not require the normalizing constant (i.e. the marginal density of the data) which may be prohibitively expensive to compute. However for most of the popular MCMC methods, such the random-walk Metropolis-Hastings (MH) algorithm, the expected computation needed to draw a nearly independent sample from the posterior grows on the order of $d^2$ (\cite{neal2011mcmc}), where $d$ is the dimensionality of the parameter vector. This `curse of dimensionality' renders the MH algorithm ineffective for high-dimensional problems.

A variant of the MH algorithm, known as Hamiltonian Monte Carlo (HMC), has gained popularity for generating samples from the posterior kernel. HMC uses  information about the gradient of the logarithm of the posterior kernel to obtain samples from a proposal distribution. These informed proposals avoid much of the random-walk behavior that plagues Metropolis-Hastings to provide much faster mixing times, as the number of steps needed is only on the order of $d^{5/4}$ (\cite{neal2011mcmc}). Though HMC's mixing properties are much better than random-walk MH, HMC comes with a much higher cost to generate samples from the proposal distribution. Using the popular `leap-frog' algorithm to perform the simulation of Hamiltonian dynamics, each sample requires evaluating the gradient of the log-posterior kernel with respect to each parameter being sampled some number $L$ times, where reasonable values for $L$ can range anywhere from 10 to 10,000 depending on the complexity of the statistical model. This can be a severe computational burden when the dimensionality of the parameter is very high. 

Herein we propose simple modifications to alleviate the expensive portions of HMC. We formulate each task in terms of simple matrix or element-wise operations that can be readily carried out on a graphics processing unit (GPU). We also propose the use of a set of persistent GPU objects to avoid the cost involved with transferring data between main and GPU memory. While these changes may seem like only a minor set of modifications, the resulting speedup is shown to be significant. In addition, with simplicity comes generality, so the techniques illustrated here should be applicable to a wide range of Bayesian models based on HMC sampling.

In the remainder of this section we will briefly discuss the basics of HMC and outline a framework for implementation using the Python programming language. In section 2, we explicitly show the proposed representation for multinomial regression. In section 3 we demonstrate the effectiveness first using a synthetic  dataset and next using a real, multi-class classification problem. We compare timing results with a popular penalized regression package in R.

\subsection{Hamiltonian Monte Caro: A Brief Overview}
The primary goals of the Bayesian inference are to obtain probabilistic inference about a parameter of interest and make predictive inference based on a given set of data. The model consists of a sampling density ($f(x|\theta)$) that generates data ($X$) conditional on the parameter ($\theta$) and a prior density ($\pi(\theta)$). Bayesian inference is based on the posterior density, $p(\theta|x)$ given by the conditional density of $\theta$ given $X=x$:
\[
p(\theta|x)=\frac{f(x|\theta)\pi(\theta)}{m(x)}
\]
where $m(x)=\int f(x|\theta)\pi(\theta)\;d\theta$ denotes the marginal density of data at $X=x$. It is well known, even for most common models, the marginal density $m(x)$ can not be computed analytically or by means of (deterministic) numerical methods especially when the dimension $d$ of $\theta$ exceeds 5 (e.g. see Chapter 5 of \cite{davis2007methods}). In such cases, Monte Carlo methods are often used which avoids the use of normalizing constant $m(x)$ by using the ratios of the posterior density evaluated at two different values of $\theta$. In particular, notice that:
\[
\frac{p(\theta|x)}{p(\theta^\prime|x)} = \frac{f(x|\theta)}{f(x|\theta^\prime)}  \cdot{ \frac{\pi(\theta)}{\pi(\theta^\prime)}} = \frac{L(\theta|x)}{L(\theta^\prime|x)} \cdot{\frac{k(\theta)}{k(\theta^\prime)}},
\]
where $L(\theta|x)=c(x)f(x|\theta)$ and $k(\theta)=c\pi(\theta)$ denote the likelihood function and prior kernel that do not involve any multiplicative components which are functions of data $x$ only. Most of the popular Monte Carlo methods, including the entire suite of MCMC methods, make use of the above property to generate samples from the posterior distribution based on only the posterior kernel $K(\theta|x)=L(\theta|x)k(\theta)$. The MH algorithm is one such approach to draw (dependent) samples from the posterior distribution using the sample path of a Markov chain whose stationary distribution is the posterior distribution (\cite{metropolis1953equation}, \cite{hastings1970monte}). Starting with an arbitrary initial value $\theta^{(0)}$, the MH algorithm proceeds iteratively by generating $\theta^{(t)}$ from the previous value $\theta^{(t-1)}$ using the following two steps:

\begin{itemize}
\item Generate a candidate value $\theta$ from a (conditional) proposal density $T(\theta|\theta^{(t)})$
\item Accept $\theta^{(t)}=\theta$ with probability $\rho(\theta|\theta^{(t-1)})$
\item If $\theta$ is not accepted,  set $\theta^{(t)}=\theta^{(t-1)}$.
\end{itemize}
The crux of the algorithm depends of the acceptance probability given by:
\[
\rho(\theta|\theta^\prime)=\min\left(1, \frac{K(\theta|X)T(\theta^\prime|\theta)}{K(\theta^\prime|X)T(\theta|\theta^\prime)}\right)
\]
It can be shown that the above scheme generates a Markov chain $\{\theta^{(t)};\;t=0,1,2,\ldots\}$ whose stationary distribution is the posterior distribution $p(\theta|x)$ under very mild regularity conditions on the proposal density $T(\theta|\theta^\prime)$ (\cite{tierney1994markov}). Hamitonian Monte Carlo (HMC) (\cite{duane1987hybrid}) modifies this basic algorithm by simulating Hamiltonian dynamics to propose new states. HMC moves in directions of high posterior density by taking steps guided by the gradient of the log-posterior with respect to $\theta$, similar in spirit to \emph{gradient descent} algorithms used in optimization problems. Here we present briefly the technical aspects of HMC, but for a more comprehensive treatment, please see \cite{neal2011mcmc}.

HMC introduces auxiliary `momentum' variables that are used to simulate and update the 'position' variables. The position variables represent the parameter vector we are interested in sampling. These two sets of variables are updated according to Hamilton's equations. In practice a discretized version, known as the 'leap-frog' method is often used to update the momentum and position vectors. Given a momentum vector at iteration $t$, $\eta^{(t)}$, and a parameter vector of interest, $\theta^{(t)}$, the update proceeds by first taking a half-step of size $\epsilon/2$ for $\eta^{(t)}$, a full step of size $\epsilon$ for $\theta^{(t)}$, and final half step for $\eta^{(t)}$:
\begin{align} 
\eta^{(t+\frac{1}{2})} &= \eta^{(t)} + \frac{\epsilon}{2} \cdot \nabla \log(K(\theta^{(t)}|X)) \label{hmc1}\\
\theta^{(t+1)} &= \theta^{(t)} + \epsilon \cdot \eta^{(t+\frac{1}{2})}\\
\eta^{(t+1)} &= \eta^{(t+\frac{1}{2})} +  \frac{\epsilon}{2}\cdot \nabla \log(K(\theta^{(t+1)}|X)) \label{hmc3}
\end{align}

Accordingly, the acceptance probability has to be modified to incorporate the momentum variables. If we could simulate Hamilton's equations perfectly, we would accept every proposal, but because the discretization introduces error, we use the following the acceptance procedure to ensure the validity of the Markov chain is intact:

\begin{align} \label{accept}
\min\left(1,\frac{\exp \left( \log(K(\theta^{(t+1)}|X)) - \eta^{(t+1)} \bullet \eta^{(t+1)} \right)}{\exp \left( \log(K(\theta^{(t)}|X)) - \eta^{(t)} \bullet \eta^{(t)} \right)} \right)
\end{align}
where $\eta^{(t+1)} \bullet \eta^{(t+1)}$ is the dot-product of the vector $\eta^{(t+1)}$ with itself. The procedure outlined in (\ref{hmc1})-(\ref{hmc3}) is one `leap-frog' update. It is usually desirable to attempt large moves from the current state, so often $L$ leap-frog updates are done per iteration, where $L$ now becomes a tuning parameter of the algorithm, as is the step size $\epsilon$. HMC thus requires $L$ evaluations of the gradient for the generation of every proposal state, followed by an evaluation of the log-posterior kernel to decide if the proposal should be accepted or rejected. The choice of the stepsize $\epsilon$ and the trajectory length $L$ depends on the particular statistical model and judicious choices must be made to acheive good mixing of the Markov chain. Preliminary runs and trace plots are often used to select these crucial tuning parameters, while automatic selection remains the central challenge to `turnkey' HMC methods. Work continues to be done on fully automated HMC with recent efforts (\cite{hoffman2011no}) achieving good success. However, all HMC methods remain computationally expensive relative to their non-gradient based MCMC counterparts and, as we illustrate through examples, this disadvantage can be reduced if we leverage GPU-based computing.

\subsection{GPU Programming in Python}
The use of general purpose graphics processing units (GPGPUs) for computationally intensive tasks has seen a considerable increase in recent years due to their potential to yield impressive speed-ups for certain classes of problems. GPU computing represents a substantially different paradigm than traditional CPU-based processing which typically executes instructions serially using a single (or as is becoming more common several) processor(s). In contrast, GPUs are composed of 100s to 1000s of processing cores that are able to perform computations on blocks of data in a highly parallel fashion. However  they do not, in general, perform serial tasks well, so GPUs are maximally useful when computation can be decomposed in terms of a self-contained function (often called a `kernel')  that can be executed in parallel on sub-blocks of data (we omit using this definition of a kernel further to avoid confusion with the statistical concept of a kernel used in this paper). Frameworks to aid in program design of this single instruction, multiple data (SIMD) paradigm have seen development from hardware vendors and open-source projects. Currently, two of the most popular frameworks are Nvida's proprietary Compute Unified Device Architecture (CUDA, \cite{nvidia2008programming}) which can only be used with Nvidia hardware and OpenCL(\cite{khronos2008opencl}) which seeks to enabled open-source, platform independent GPU computing. Both frameworks provide C-like compilers that allow users to write GPU functions in a familiar syntax.

Python (\cite{sanner1999python}) is an interpreted programming language available for all major operating systems. It has a growing set of statistical libraries and features a syntax that will be familiar to R users. Python has a increasingly mature set of GPU interfaces that facilitate easy parallel programming, a feature which R presently lacks. The PyCuda library (\cite{klockner2012pycuda}) is built on top of the CUDA and platform provides many abstractions that make GPU programming much easier. It features support for GPU array objects that mirror Pythonic arrays and are  similar in nature to matrix objects in R. 

The GPU array abstraction allows for operations on and between GPU arrays without the need to explicitly manage the underlying hardware. Once a GPU array has been created, it can be used much like a normal array, except that the computation happens using the GPU as opposed to the central processing unit (CPU). Moreover, it uses Python's built in memory management to automatically free dereferenced objects, freeing the programmer from having to explicitly free unused memory. Should a problem require a custom piece of GPU code, PyCuda features a `just-in-time' compiler that can compile source modules at runtime, which provides a great deal of flexibility. Additionally, the \emph{cuda} package in the SciPy library (\cite{jones2001scipy}) is built on top of PyCuda and offers many linear algebra functions. One of the draw backs of typical GPU programming is that execution code and data must be transferred from main memory to the GPU's local memory. This transfer incurs a latency penalty, which may result in a GPU-based program taking longer to execute than a CPU implementation if the latency is large relative to the execution time. Using PyCuda and the cuda-SciPy libraries, we can easily create persistent GPU based objects and repeatedly update these objects which remain in GPU memory to avoid transfer latencies. We note that while we use Python for demonstration purposes, the approaches described here should be easily ported to any CUDA based programming environment. 

\section{Methods}
\subsection{Bayesian Multinomial Regression}
Let $x_i = <x_{i1},\ldots,x_{ip}>^T$ be a vector of predictors and $y_i = <y_{i1},\ldots,y_{iK}>^T$ be a vector of mutally exclusive indicators of class membership, where each $y_{ik} \in \{0,1\}$. We wish to classify new observations into 1-of-$K$ classes. The relationship between $x_i$ and $y_i$ is achieved through the generalized logit, often called the \emph{softmax} function: 
\begin{align} \label{softmax}
Pr(y_{ik} = 1 | x_i,\beta_k) = \psi(x_i,\beta_k) = \frac{exp(x_i^T \beta_k)}{\sum_{j=1}^{K} exp(x_i^T \beta_j)}
\end{align}
where $\beta_k$ is the the $p\times 1$ vector of regression coefficients associated with the $k$-th class. Assuming each $y_i$ follows a multinomial distribution, the log-likelihood for all $n$ observations, $l(B|X,Y)$, is:
\begin{align} \label{loglike}
l(B|X,Y) &= \sum_{i=1}^n \sum_{k=1}^K y_{ik}\log(\psi(x_i,\beta_k))
\end{align}
where $B_{p \times k}= [\beta_1, \dots , \beta_K]$ is the matrix containing the vectors of regression coefficients for all $K$ classes. We set $\beta_K=0$ to ensure identifiability of the parameters (notice that as the multinomial probabilities add up to one, we need a constraint to identify the $\beta$'s uniquely). Following the recommendations in \cite{gelman2008weakly}, we place a non-conjugate, weakly informative standard Cauchy distribution on each element of $B$. In this notation $\beta_{jk}$ indexes the single regression coefficient of $B$ associated with variable $j$ for class $k$, while $\beta_k$ refers to the $p \times 1$ vector of all regression coefficients for class $k$. The corresponding log-posterior kernel is:
\begin{align}\label{logkernel}
\log(K(B|X,Y)) =\sum_{i=1}^n \sum_{k=1}^K y_{ik}\log(\psi(x_i,\beta_k)) + \sum_{k=1}^K\sum_{j=1}^p - \log(1+\beta_{jk}^2)
\end{align}
where $X_{n \times p} = [x_1, \ldots, x_n]^T$ and $Y_{n \times k} = [y_1, \ldots , y_n]^T$. We can now rewrite the first term of the log-likelihood as $Y \star \log(\psi(XB))$, which can be computed using only linear or element-wise operations. In detail, $XB$ is carried out via matrix-matrix multiplication, followed by a row-wise softmax transformation $\psi(\cdot)$, and finally an element-wise $\log(\cdot)$ transformation. An element-wise multiplication between $Y$ and the resulting $n \times k$ matrix, $\log(\psi(XB))$, yields the log-likelihood values for all $n$ observations and $k$ classes.  Instead of looping over elements of this matrix to sum the total log-likelihood value, we can pre-multiply by $1^T_{1\times n}$ to sum over all observations and post-multiply by $1_{k\times1}$ to sum over all classes. The contribution from the prior in (\ref{logkernel}) can also be computed in similar fashion. Let $\Gamma_{p \times k}$ be the matrix containing the prior's contribution, where $\Gamma[j,k] =-\log(1 + \beta^2_{jk})$. The equivalent form of (\ref{logkernel}) that only uses linear or element-wise operations is shown below.
\begin{align} \label{logkernelGPU}
\log(K(B|X,Y)) = 1^T_n\left(Y*\log(\psi(XB))\right)1_k + 1^T_p \Gamma  1_k
\end{align}
Using similar reasoning, we can rewrite the evaluation of the gradient in terms of matrix and element-wise operations. The result is shown below, please see the appendix for the full derivation in addition to discussion on maintaining identifiability during the gradient updates.

\begin{align} \label{gradient}
\left[\frac{\partial \log(K(B|X,Y))}{\partial B}\right]_{p \times k} &= X^T \left(Y - \psi(XB)\right) + \Gamma
\end{align}
For this statistical model, we propose the use persistent PyCuda GPU arrays for $X,Y,B,\eta,$ and $\left[\frac{\partial \log(K(B|X,Y))}{\partial \beta}\right]$, allowing us to perform the entire HMC simulation in GPU memory. without having to move objects from main memory into the GPU's local memory.

\section{Results}
\subsection{Computational Efficiency Analysis}
We begin by first presenting some timing results for both the CPU and GPU implementations that will serve as a baseline for comparison. All simulations were written in Python and executed on a computing cluster node with 2 Xeon E5-2670 processors, 256GB of RAM, and a NVIDIA Tesla K20 GPU with 2500 CUDA cores and 5GB of RAM using NVIDIA's CUDA 5.0 SDK. The CPU version was implemented using the numpy library and the GPU version was implemented using the previously mentioned PyCuda and CUDA scikit for SciPy. It should be noted that the numpy library for linear algebra in Python is a high-level wrapper for the BLAS library (\cite{lawson1979basic}) and will, in general, be comparable in speed to a pure C implementation for linear algebra tasks. The CPU and GPU implementations used identical representations, with the only differences being due to differences required by numpy and PyCuda. However, any differences in execution speed should be largely attributable to speeds offered by the hardware the code was executed on (i.e. CPU vs GPU). For all computations, 32-bit floating point numbers were used in both GPU and CPU implementations. 

\begin{figure}[h!]
\centering
	\subfloat{\includegraphics[width=3.5in]{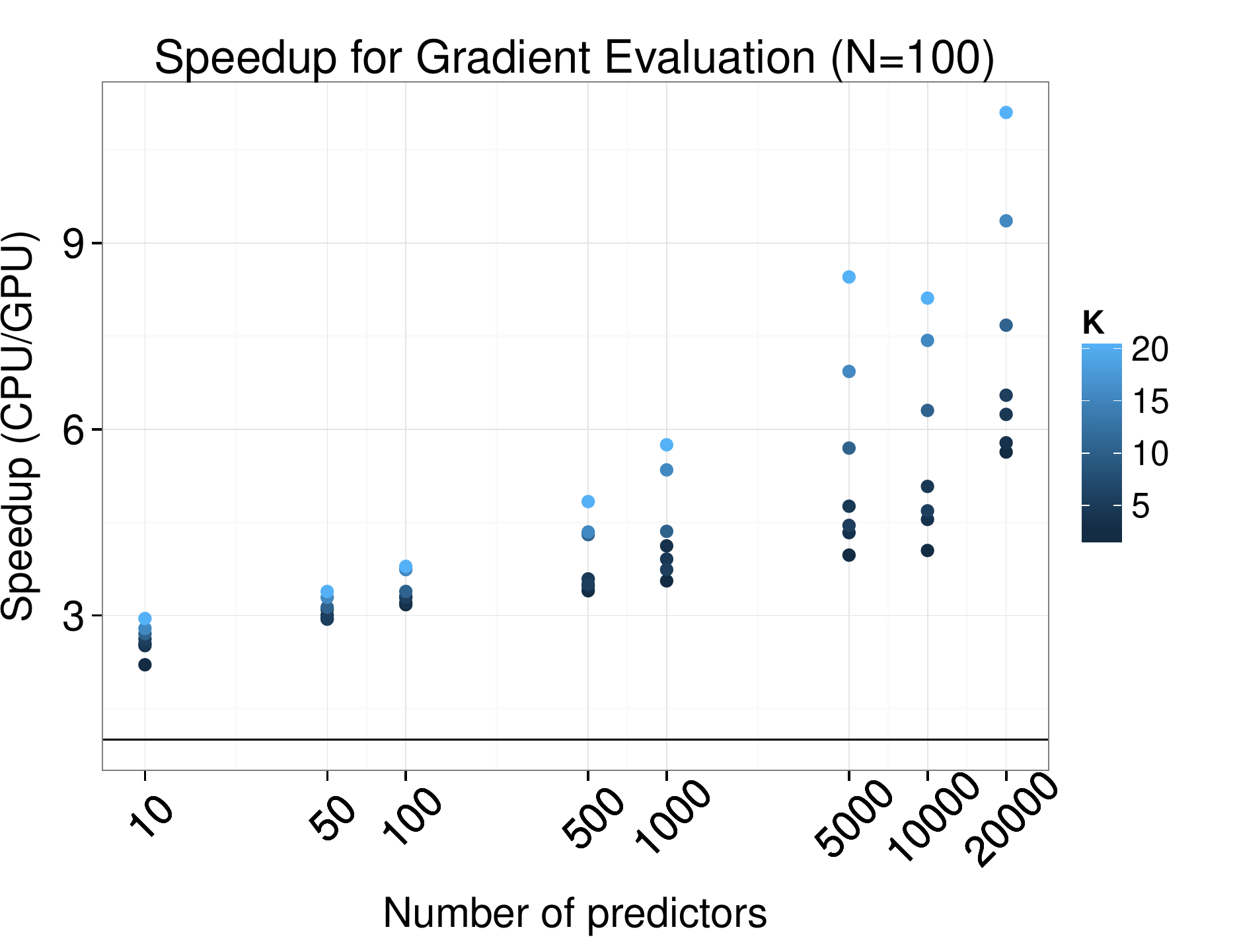}}
	\subfloat{\includegraphics[width=3.5in]{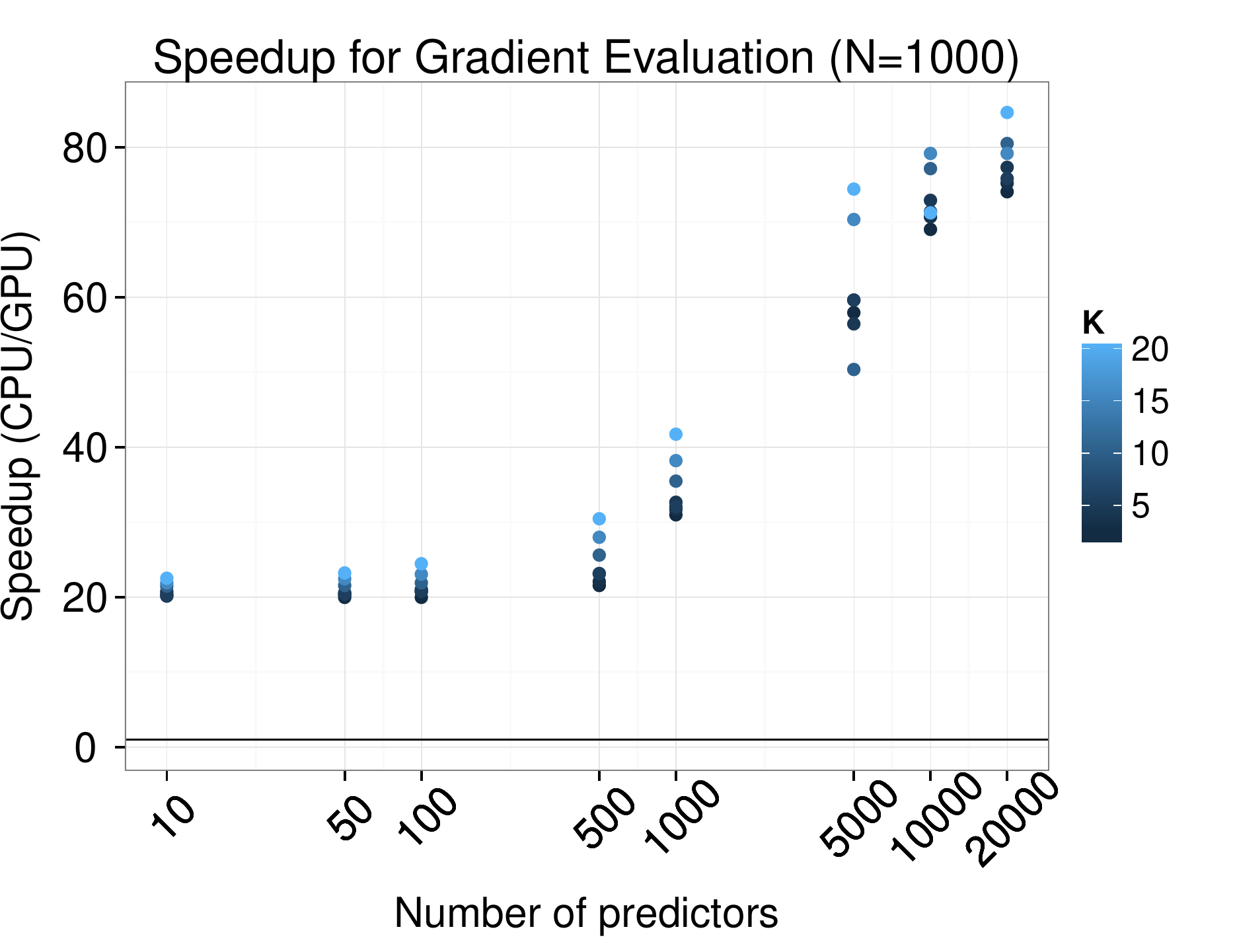}} \quad
	\subfloat{\includegraphics[width=3.5in]{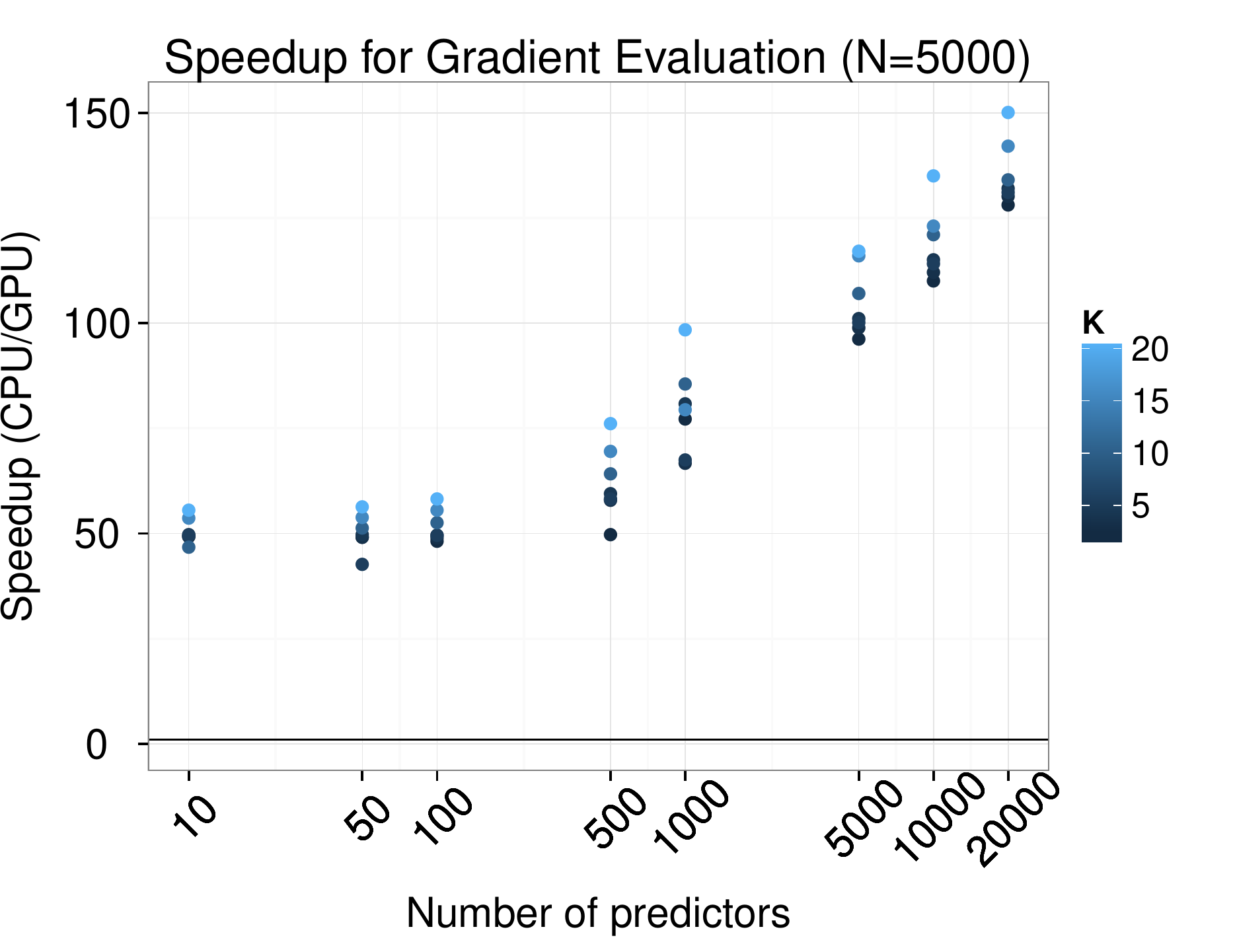}}
	\subfloat{\includegraphics[width=3.5in]{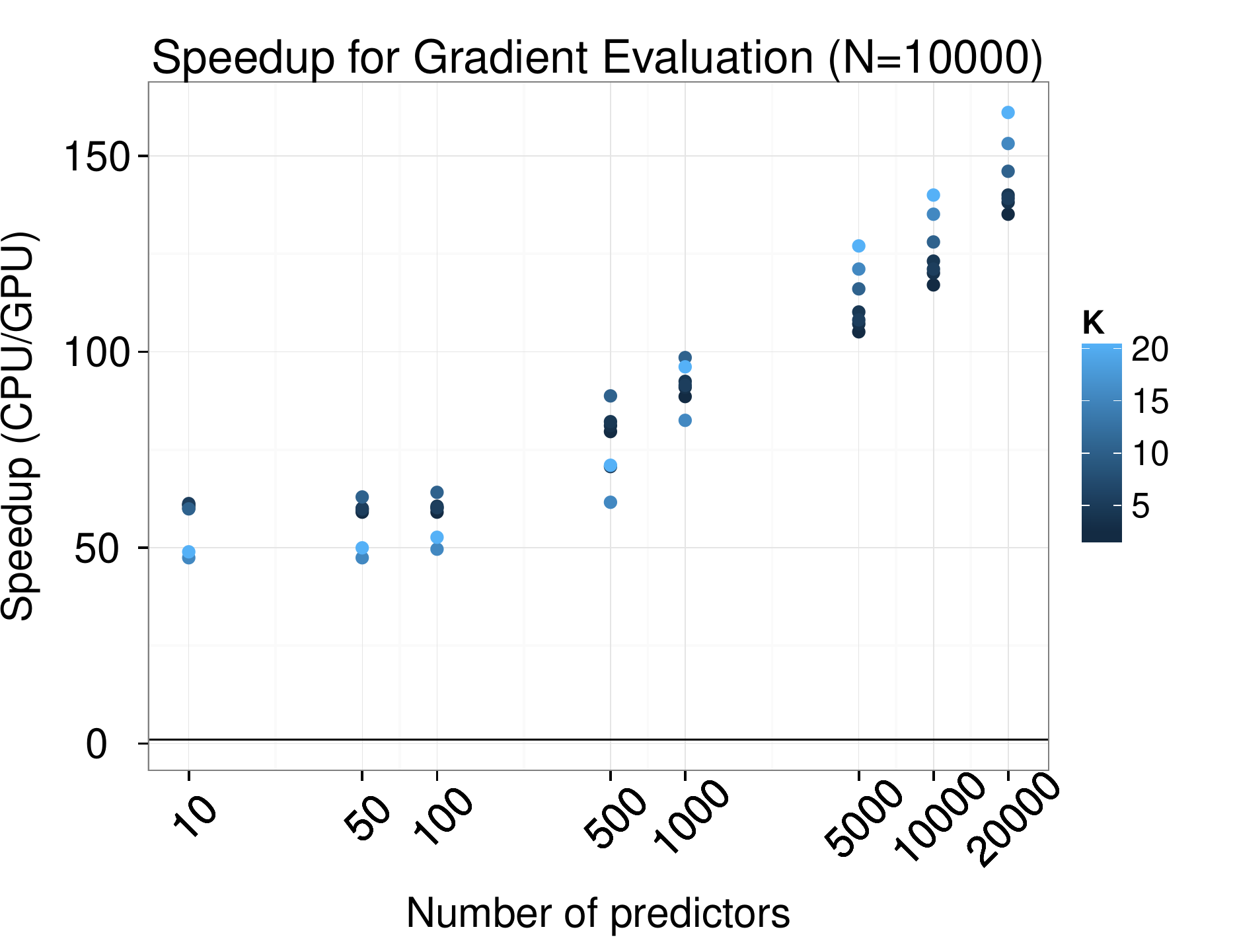}} \quad
		\caption{Average timing results for one gradient evaluation for various combinations of sample size, parameter dimension and number of classes. Each dot represents one such combination. Each panel represents a different sample size, the x-axis is the number of predictors, and the color of the dot represents the number of classes (lighter colors correspond to larger values).The y-axis is the time taken for the CPU divided by the time taken for the GPU. The black horizontal line is at 1 - values below this line are faster for the CPU and values above are faster on the GPU.} \label{fig1}
\end{figure}

\begin{figure}[h!]
\centering
	\subfloat{\includegraphics[width=3.5in]{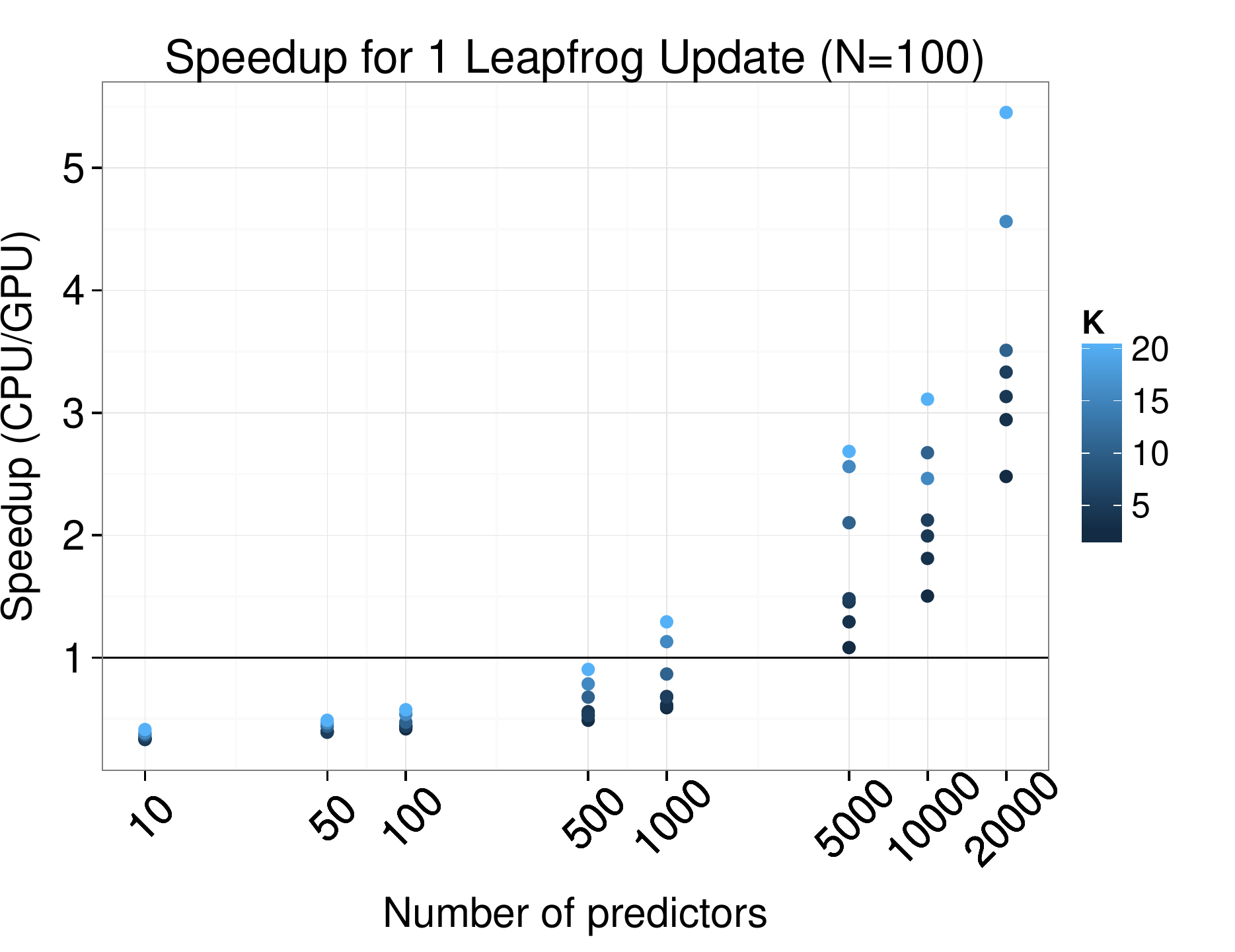}}
	\subfloat{\includegraphics[width=3.5in]{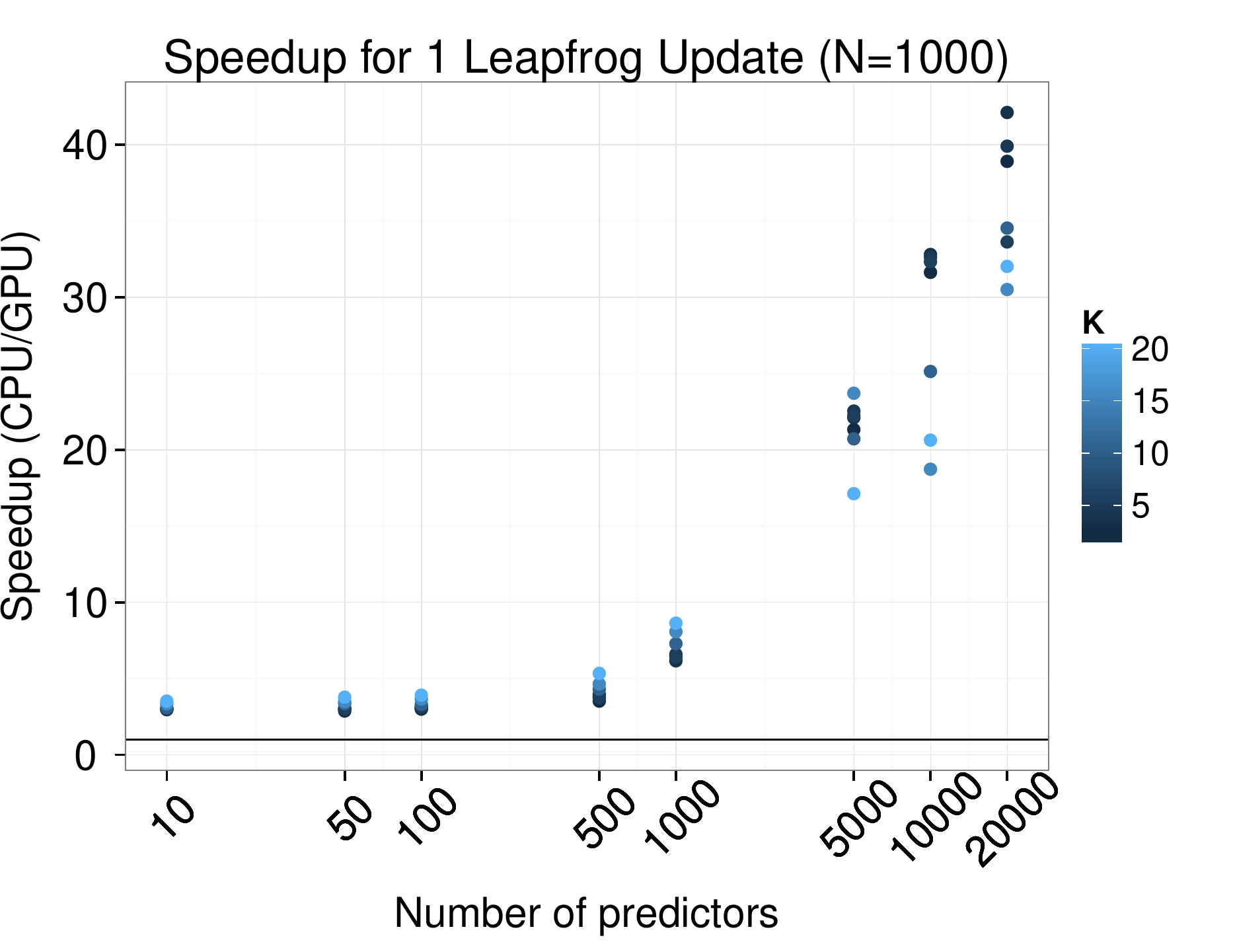}} \quad
	\subfloat{\includegraphics[width=3.5in]{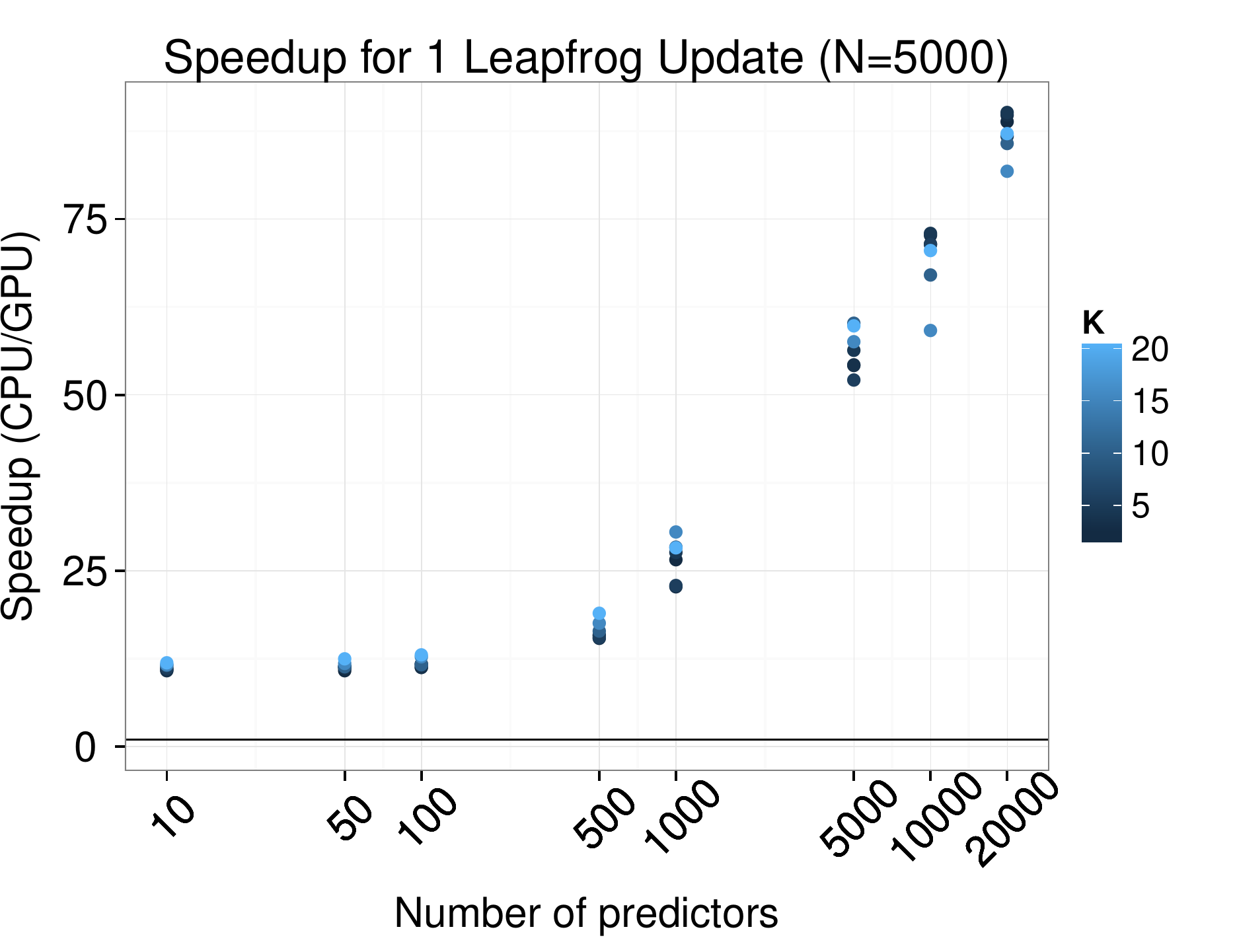}}
	\subfloat{\includegraphics[width=3.5in]{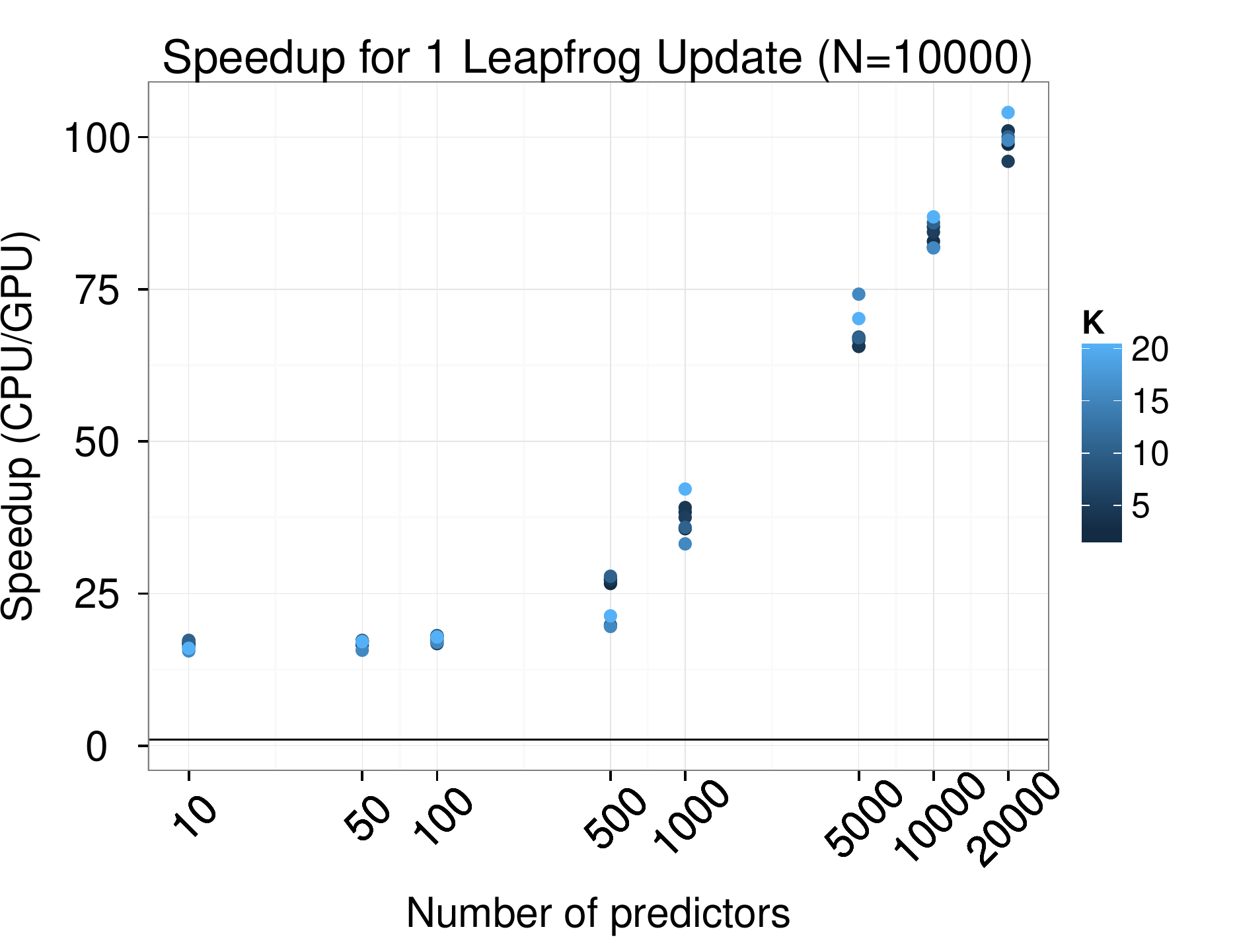}} \quad
		\caption{Average time required for one 1 leap-frog update for various combinations of sample size, parameter dimension, and number of classes. Each dot represents one such combination. Each panel represents a different sample size, the x-axis is the number of predictors, and the color of the dot represents the number of classes (lighter colors correspond to larger values).The y-axis is the time taken for the CPU divided by the time taken for the GPU. The black horizontal line is at 1 - values below this line are faster for the CPU and values above are faster on the GPU.} \label{fig2}
\end{figure}

For several values of sample size ($N$), number of classes ($K$), and dimension ($p$) we performed a single gradient evaluation and a separate, single leap-frog update five times and recorded the the average time taken. Each value of $X$ and $B$ was drawn from a $N(0,10^{-3})$. $Y$ generated according to (\ref{softmax}), given the values of $X$ and $B$. We swept each value shown in Table \ref{tab1}, for a total of 224 parameter combinations. The results of the simulations are summarized in Figures \ref{fig1} and \ref{fig2}.

\begin{table}[h!]
\centering
\begin{tabular}{|l|c|c|c|}
\hline
\textbf{Parameter} & N & K & p\\
\hline
\textbf{Values Swept} & $100,1000,5000,1000$ & $2,3,4,5,10,15,20$ & $10,50,100,500,1000,5000,10000,20000$\\
\hline
\end{tabular}
\caption{Parameter values used in timing evaluations.}
\label{tab1}
\end{table}

Figures \ref{fig1} and \ref{fig2} show that the computational core of the HMC algorithm (the gradient evaluation and leap-frog update) benefit greatly from the use the GPU. For large problems the GPU is 50-100 times faster than the corresponding CPU version. Only for certain instances of the smallest problem size ($N = 100$) was the GPU slower than the CPU. However, even for small sample sizes many cases were still several times faster on the GPU. Though larger problems benefit proportionally more from the parallelism offered by the GPU, problems of more modest size may still benefit from this approach.

\subsection{Handwritten Digit Recognition}
In this section, we apply the formulation of Section 2 to a real dataset. The MNIST handwritten digit dataset (\cite{lecun1995comparison}) is a popular benchmark in computer vision and machine learning. Each of the 70,0000 non-sparse observations represents the pixel intensity from a 28x28 image of a digit between 0 and 9. The images have been preprocessed, centered, and normalized such that they all are of comparable intensities. Using this dataset, we performed multinomial regression to classify each observation into one of the 10 possible classes. We used a `flattened' representation such that the 28x28 square becomes one vector with 784 features. We used the first 60,000 observations to build the model and withheld the final 10,000 as a test set. Thus our data is a predictor matrix that includes an intercept, $X_{60,000x785}$ and a response matrix, $Y_{60,0000x10}$ and a coefficient matrix $\beta_{785x10}$. We used the model outlined in Section 2 and performed HMC with $L = 100$ and $\epsilon = 10^{-4}$ during the burn in phase and $\epsilon = 7*10^{-5}$ during the sampling phase. The values of $L$ and $\epsilon$ were selected to achieve an acceptance rate between 0.65 and 0.75 for the sampling phase in accordance with the suggestions in \cite{neal2011mcmc}. In order to get the chain moving during the burnin phase, we adopt an \emph{annealed} acceptance strategy. Let $\alpha$ be the usual acceptance probability in (\ref{accept}), we modify the procedure such that at iteration $t$ we accept a proposal instead with probability $\alpha^{1/T_{(t)}}$, where $T_{(t)}$ is set according to an \emph{annealing schedule}, $T_{(t)} = \max(1,r*T_{(t-1)})$, for some $r < 1$. This procedure helps the simulation get started and lessens the chance of getting stuck in a local minima early in the simulation. We do not anneal during the sampling phase, so the stationary distribution of the Markov chain remains the one defined by (\ref{logkernel}). We allowed the chain to burn in for 100 iterations with an initial $T_{(0)} = 10^3$ and $r = 0.9$ before beginning the sampling phase. For predicting class membership, we predicted the class that had the highest average softmax value in the posterior samples.

For comparison, we also fit a penalized multinomial regression model under an $L_1$ penalty, i.e. the Lasso penalty. This model is well known to be equivalent to the \emph{maximum a posteriori} (MAP) estimate of a Bayesian posterior under a Laplace prior. We used the \emph{glmnet} (\cite{friedman2010regularization}) package in R, which is uses one of the most efficient methods available for this type of model. In \cite{friedman2010regularization} the authors compared \emph{glmnet} to a popular approximate Bayesian software package BBR (\cite{genkin2007large}) that also produces a MAP estimate for each parameter. At that time BBR was one of the fastest approaches available and Friedman et. al found that glmnet was significantly faster on most datasets they analyzed. This indicates that \emph{glmnet} should serve as a good baseline for timing performance evaluation. Note also that the approach offered here does not provide a MAP estimate, but instead provides samples from the \emph{full} posterior distribution.

In our comparison the shrinkage parameter ($\lambda$) in \emph{glmnet} was selected via 5-fold cross-validation. Using the parallel processing option in \emph{glmnet}, we were able to fit each fold simultaneously on separate CPU cores, making the time required for 5-fold CV equivalent to fitting a single model.  The number of $\lambda$ values tested during the cross-validation has a large impact on the time needed to build the model, so we first used the default of 100 values for $\lambda$ and then fit using larger grids of 500 and 1000. The results are summarized in the Table \ref{tab2}:

\begin{table}[h!]
\centering
\begin{tabular}{|l|c|c|c|c|c|}
\hline
 & \multicolumn{2}{|c|}{\textbf{Bayesian-HMC}} &  \multicolumn{3}{|c|}{\textbf{glmnet}} \\
\hline
Tuning Parameter & 100 samples & 500 samples & $\# \lambda = 100 $ (default) & $\# \lambda = 500$  & $\# \lambda = 1000$  \\
\hline
Train Accuracy & 0.928 & 0.932 & 0.922 & 0.932 & 0.932 \\
\hline
Test Accuracy  & 0.92 & 0.921 & 0.920 & 0.926 & 0.926\\
\hline
Time (hr:min:sec)  & 0:3:54 & 0:11:42 & 6:43:59 & 2:25:12 \tablefootnote{A larger grid of values resulting in a shorter execution time than a smaller grid was confirmed as normal behavior. Personal correspondence with \emph{glmnet} package maintainer.} & 3:41:3$^{1}$\\
\hline
\end{tabular}
\caption{Results for MNIST dataset. The Bayesian-HMC method always burned in for 100 iterations and used $L=100$ and $\epsilon=7*10^{-5}$ for HMC sampling after the burn in phase.}
\label{tab2}
\end{table}

\section{Discussion}
We have outlined a general method for performing HMC-based Bayesian analyses using GPUs. Our results show that GPUs can greatly speed up the expensive parts of HMC, often reducing the time needed for the core components of HMC by more the 100-fold for large problems. This is a considerable amount time saved and not only reduces tedium experienced by the analyst waiting for a simulation to finish, but also opens Bayesian analyses to a larger class of problems. This framework should easily extend to more complex models, including hierarchical models, as the gradient calculations can be propagated from the highest levels down to the lowest levels using matrix multiplication. There many additional libraries available in Python not explored here that can aid in running these types of models on a GPU. In particular, Theano (\cite{bergstra2010theano}) and PyAutoDiff can perform automatic differentiation and gradient evaluations using GPUs. This type of framework would free a user from having to calculate the gradient by hand and program the corresponding GPU-friendly representation.  HMC-based sampling with automatic differentiation is currently available in the stand alone software package known as  Stan (\cite{stan-software:2013}), which is growing quickly in popularity. Stan  uses a much more sophisticated sampling scheme than the one presented here, namely it uses the `No-U-Turn' (NUTS) sampler that adaptively tunes the step size $\epsilon$ and the trajectory length $L$. We explored the use of Stan for comparison in this study, because it compiles a BUGS like model specification into fast C++ code. This is an exciting and ongoing project, but unfortunately at the time of this writing, some of the computational infrastructure needed by the models in this study were not fully optimized \footnote{Personal correspondence with Stan development team}, so we omitted Stan from the comparisons shown here.

In comparison to a popular penalized regression approach on the large MNIST dataset, we were able to complete a fully Bayesian analysis in a fraction of the time, while achieving nearly identical accuracy results. Conceivably we could have used this speedup to explore more complex models and prior structures to possibly achieve greater accuracy, but since timing was of primary interest and because of the connection between penalized regression and Bayesian models, we did not explore more sophisticated models. These results are not comprehensive evaluation of timings between these two methods, but is only meant to highlight that it is possible to do a fully Bayesian analysis in relatively short order on a large problem.

In conclusion, it seems that in the future all types of analyses must evolve to cope with the burden imposed by ever increasing amounts data. The framework presented here has the potential to enable the use of fully Bayesian approaches on these ever larger datasets. Moreover, on board GPU RAM will continue to increase according to Moore's law, allowing for ever larger simulations to be performed completely in GPU memory. However, before concluding we pause for a moment of truth-in-advertising. The type of speedup offered here is not limited to Bayesian approaches, but can be used in general by any method where likelihood and gradient evaluations can be expressed in terms of linear algebra operations such as gradient descent, Gauss-Newton, and many others. HMC is only one such method that is highly amenable to GPU speedup. However, the Bayesian approach is very attractive due to the coherent model regularization as well as the quantification of parameter uncertainty offered by the posterior distribution. Corresponding estimates of parameter uncertainty are currently very difficult for the penalized regression methods, so we find the Bayesian framework very appealing in high-dimensional settings. We hope the results shown in this study will encourage further development in the use of GPUs for high-dimensional statistical problems.

\section{Appendix}
Code from this paper is available at \url{https://github.com/beamandrew/HMC_GPU}
\subsection{Gradient derivations}
We need to compute the gradient of the log-kernel with respect to each $\beta_{jk}$. Application of the chain rule to (\ref{logkernel}) and differentiating with respect to $\beta_{jk}$ yields:

\begin{align}
\frac{\partial \log(K(B|x_i,y_i))  }{\partial \beta_{jk}} &= \frac{\partial }{\partial \log(\psi(x_i,\beta_k))}\frac{\partial \log(\psi(x_i,\beta_k))}{\partial \beta_{jk}}\left[\sum_{c=1}^K y_{ic}*\log(\psi(x_i\beta_c))\right] + \frac{\partial }{\partial \beta_{jk}} \left[-\log(1 + \beta_{jk}^2)\right]\\
\frac{\partial l\log(K(B|x_i,y_i))}{\partial \beta_{jk}} &= y_{ik}\frac{\partial \log(\phi(x_i,\beta_k))}{\partial \beta_{jk}} \left[\log(\psi(x_i,\beta_k))\right] + \frac{2*\beta_{jk}}{1 + \beta_{jk}^2}\\
\frac{\partial \log(K(B|x_i,y_i))}{\partial \beta_{jk}} &= y_{ik}- \psi(x_i,\beta_k) + \frac{2*\beta_{jk}}{1 + \beta_{jk}^2} \label{lastline}
\end{align}
where (\ref{lastline}) follows from the fact that $\frac{\partial \log(\psi(x_i,\beta_j))}{\partial \beta_{jk}} \log(\psi(x_i,\beta_k) = 1 - \log(\psi(x_i,\beta_k))$ if $y_{ik} = 1$ and $\frac{\partial \log(\psi(x_i,\beta_k))}{\partial \beta_{jk}}\log(\psi(x_i,\beta_k) = -\log(\psi(x_i,\beta_k))$ if $y_{ik} = 0$.
We need sum over all $n$ observations to obtain the full gradient for $\beta_{jk}$, but instead of summing over all cases in a loop, we pre-multiply the vector of differences, $Y_{nxk} - \psi(X_{nxp}B_{pxk})$, by $X^T_{pxn}$, yielding the matrix of partial derivatives:
\begin{align} \label{fullgrad}
\left[\frac{\partial  \log(K(\beta|X,Y))}{\partial \beta}\right]_{pxk} &= X^T \left(Y - \psi(X\beta)\right) + \Gamma
\end{align}
where the $\Gamma$ matrix is now the $p \times k$ matrix with $\Gamma[j,k] = \frac{2*\beta_{jk}}{1+\beta^2_{jk}}$. Note that the equation (\ref{fullgrad}) does not take into account the identifiability constraint, $\beta_K = 0$. In order to maintain identifiability, we must first set $\beta_K = 0$ and ensure that the gradient for $\beta_K$ is uniformly 0 (as well as the momentum component for $\beta_K$ used in the HMC procedure). The way in which we do this is context dependent. If we are operating in a programming environment that allows for sub-array access, also known as \emph{slicing}, we can simply update the sub-array of the gradient matrix containing only the unconstrained coefficients. R and Python both support this kind of memory access, however there may be some penalty involved in accessing the underlying sub-array. To evaluate the best-case scenario for CPU-based code, we maintain a \emph{separate} $(p-1) \times k$ matrix for the gradient and leap-frog update timings in Section 3.1. This would correspond to the ability to access the sub-array for the unconstrained coefficients with no penalty. This approach allows us to put a lower-bound on the speed-up offered by the GPU.

\begin{figure}[H]
\centering
	\subfloat{\includegraphics[width=3.5in]{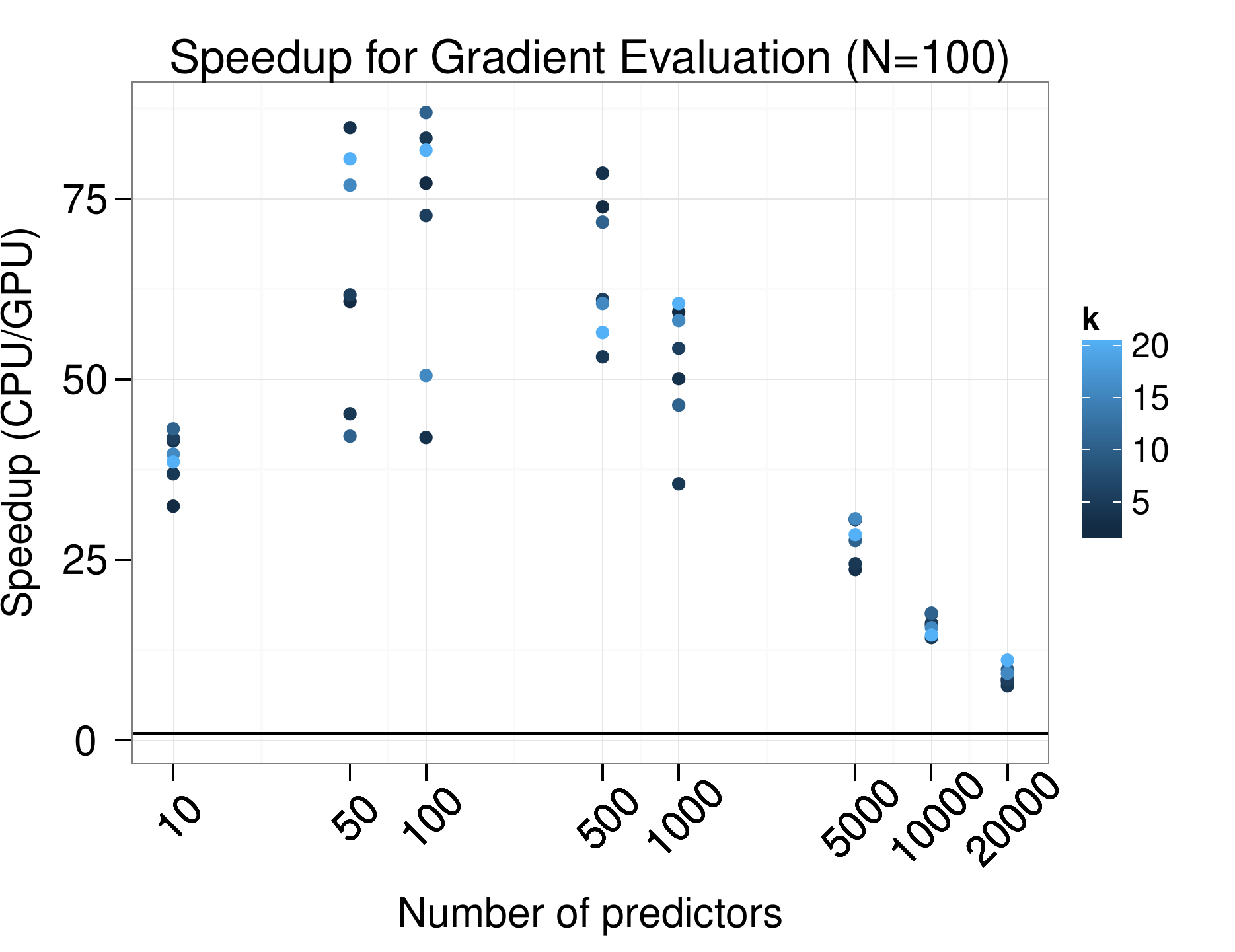}}
	\subfloat{\includegraphics[width=3.5in]{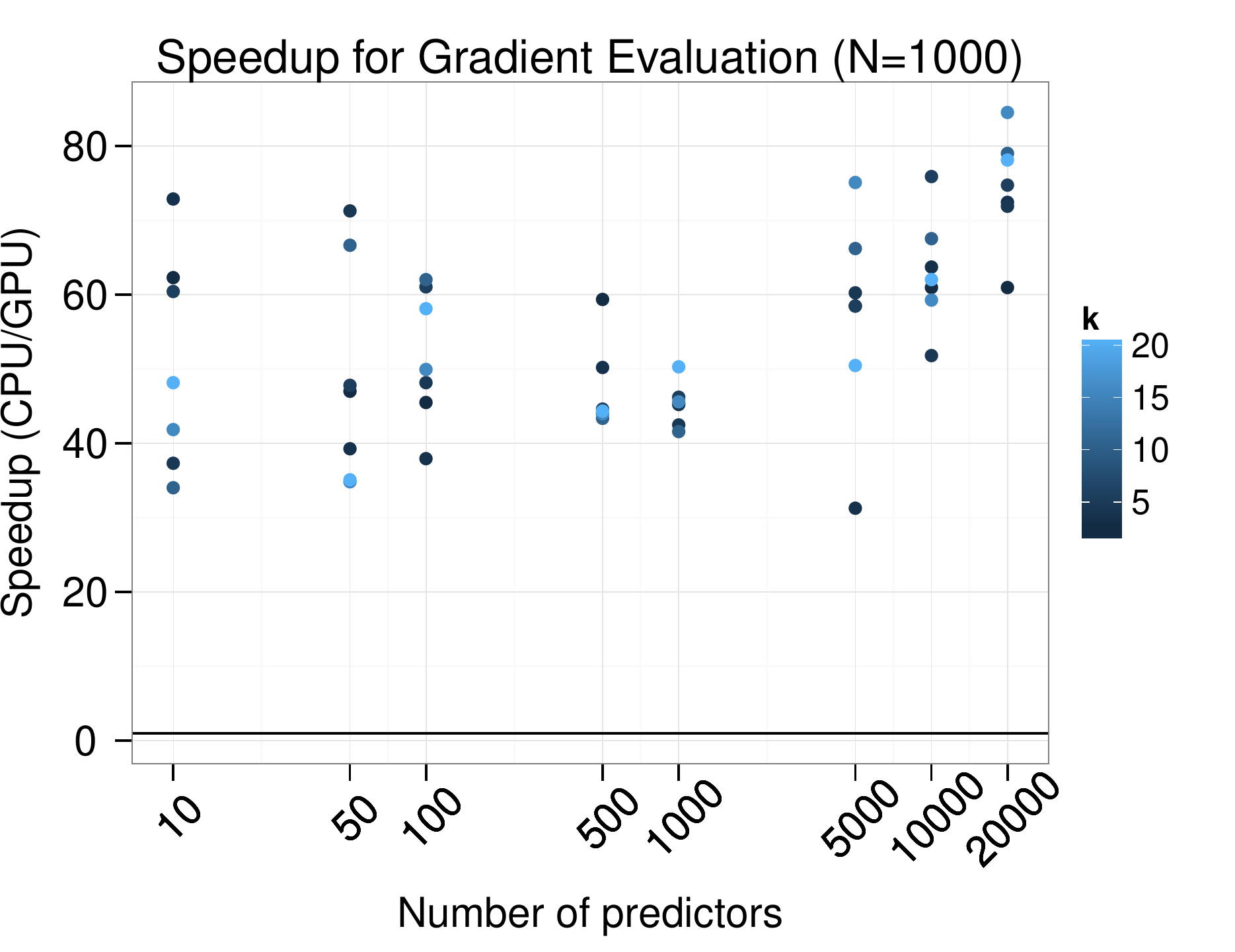}} \quad
	\subfloat{\includegraphics[width=3.5in]{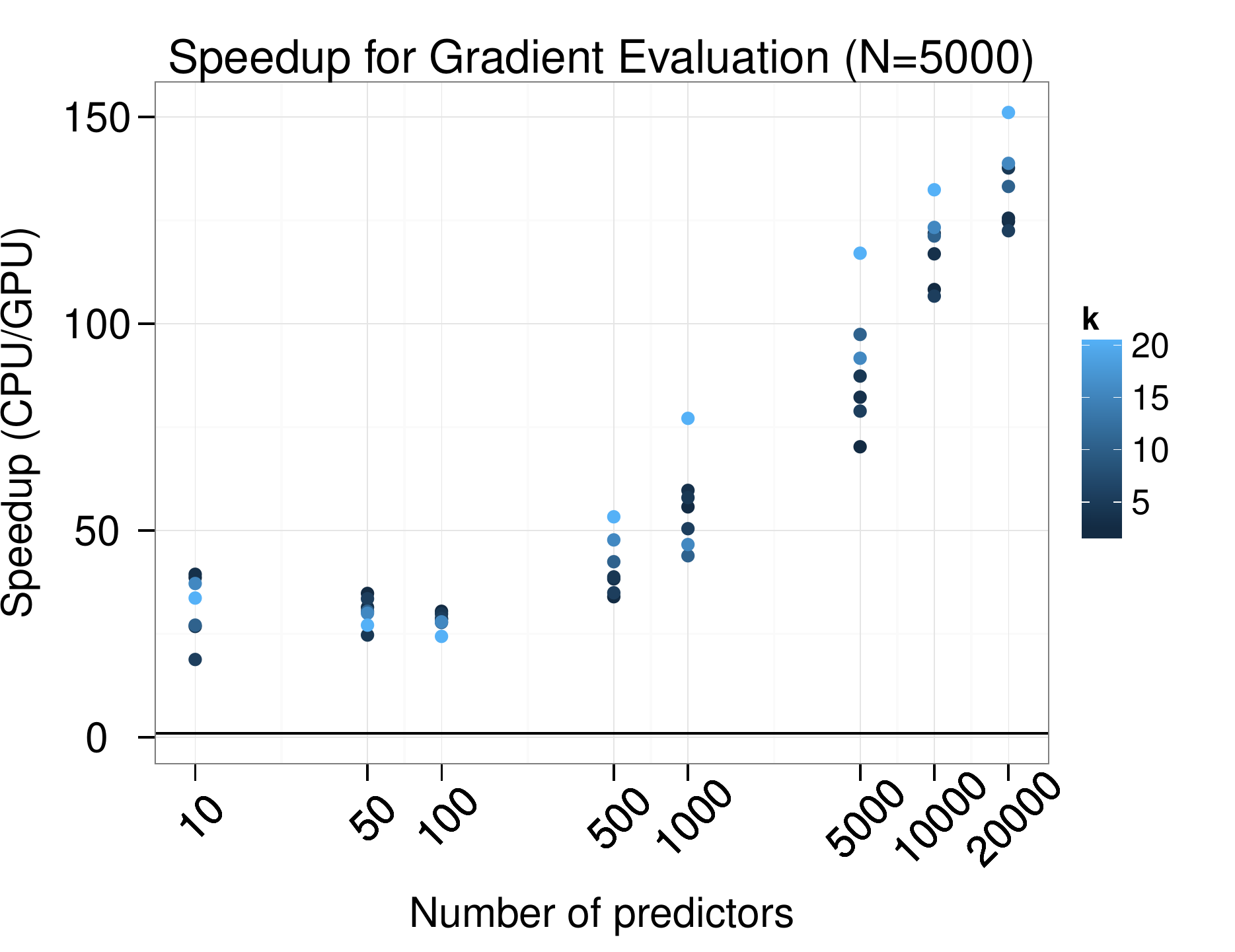}}
	\subfloat{\includegraphics[width=3.5in]{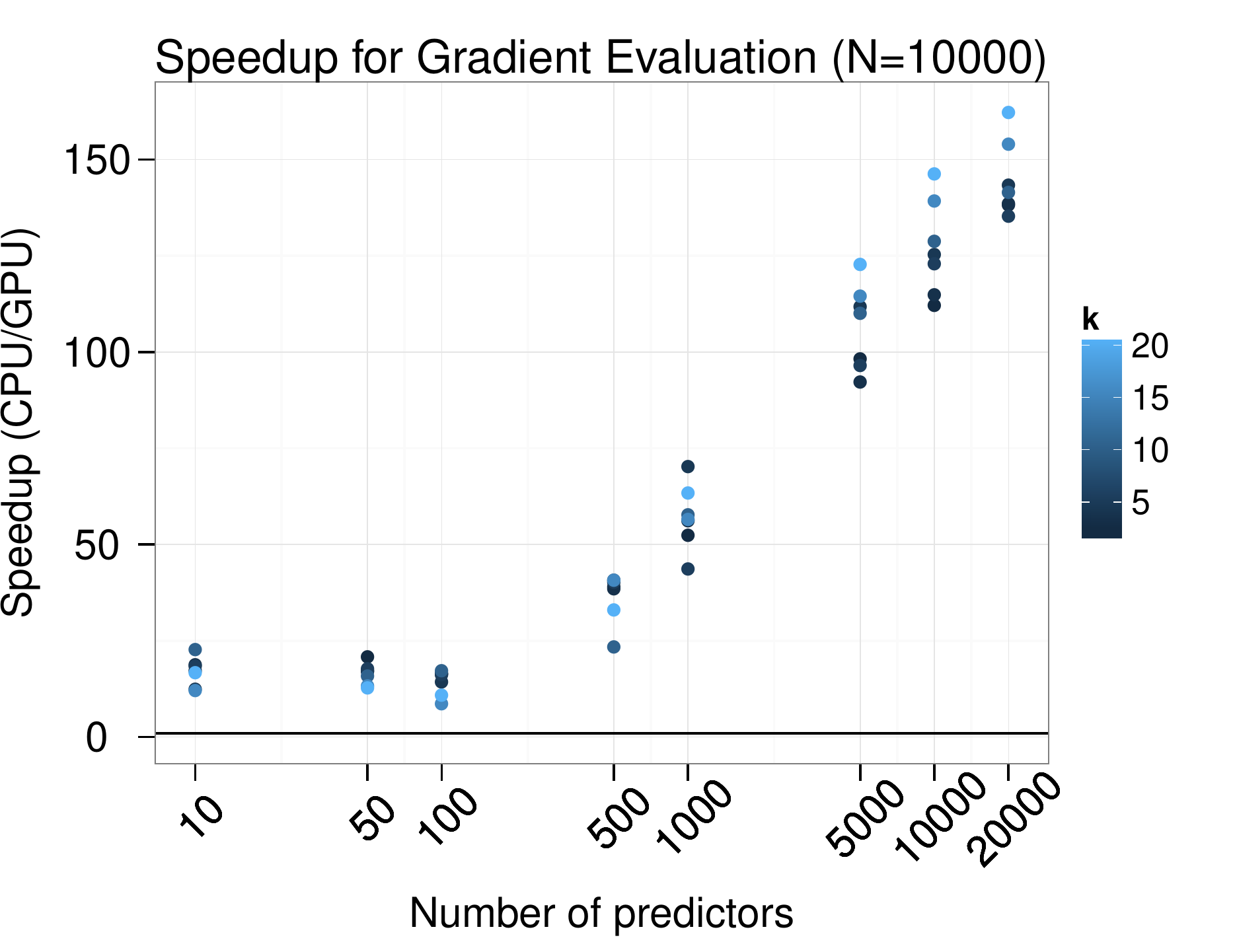}} \quad
		\caption{Timing results for gradient evaluation when identifiability is not enforced.} \label{fig3}
\end{figure}

\begin{figure}[H]
\centering
	\subfloat{\includegraphics[width=3.5in]{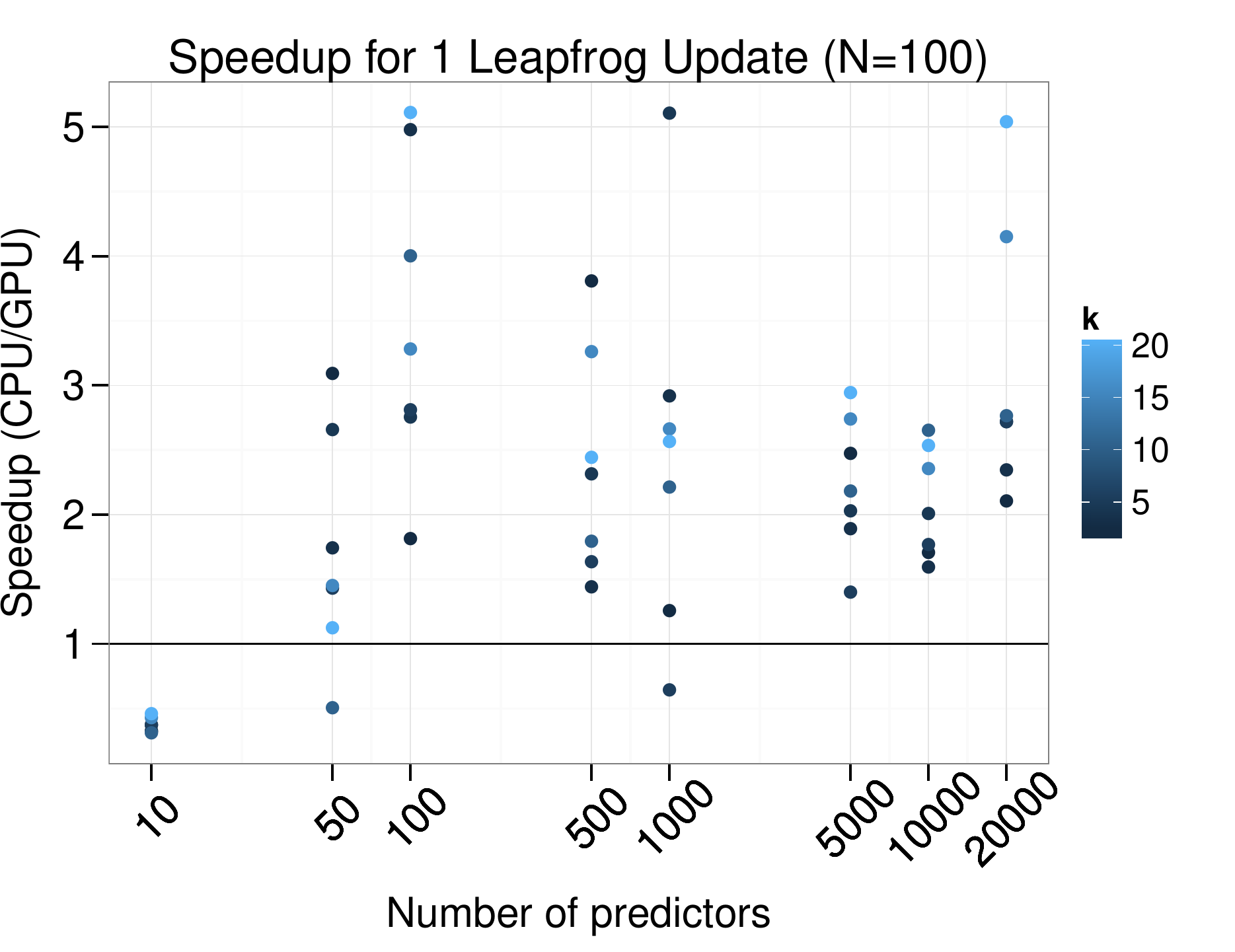}}
	\subfloat{\includegraphics[width=3.5in]{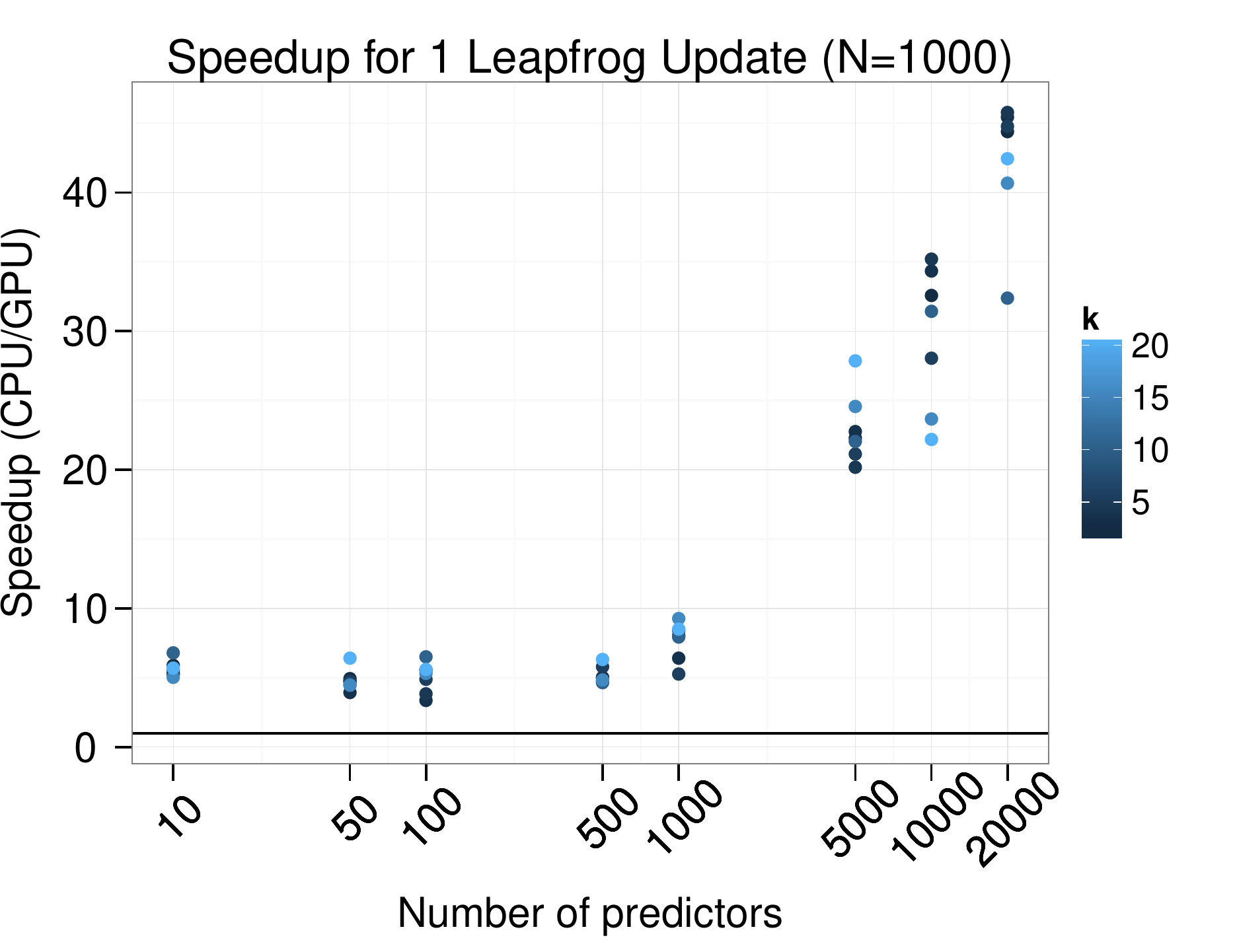}} \quad
	\subfloat{\includegraphics[width=3.5in]{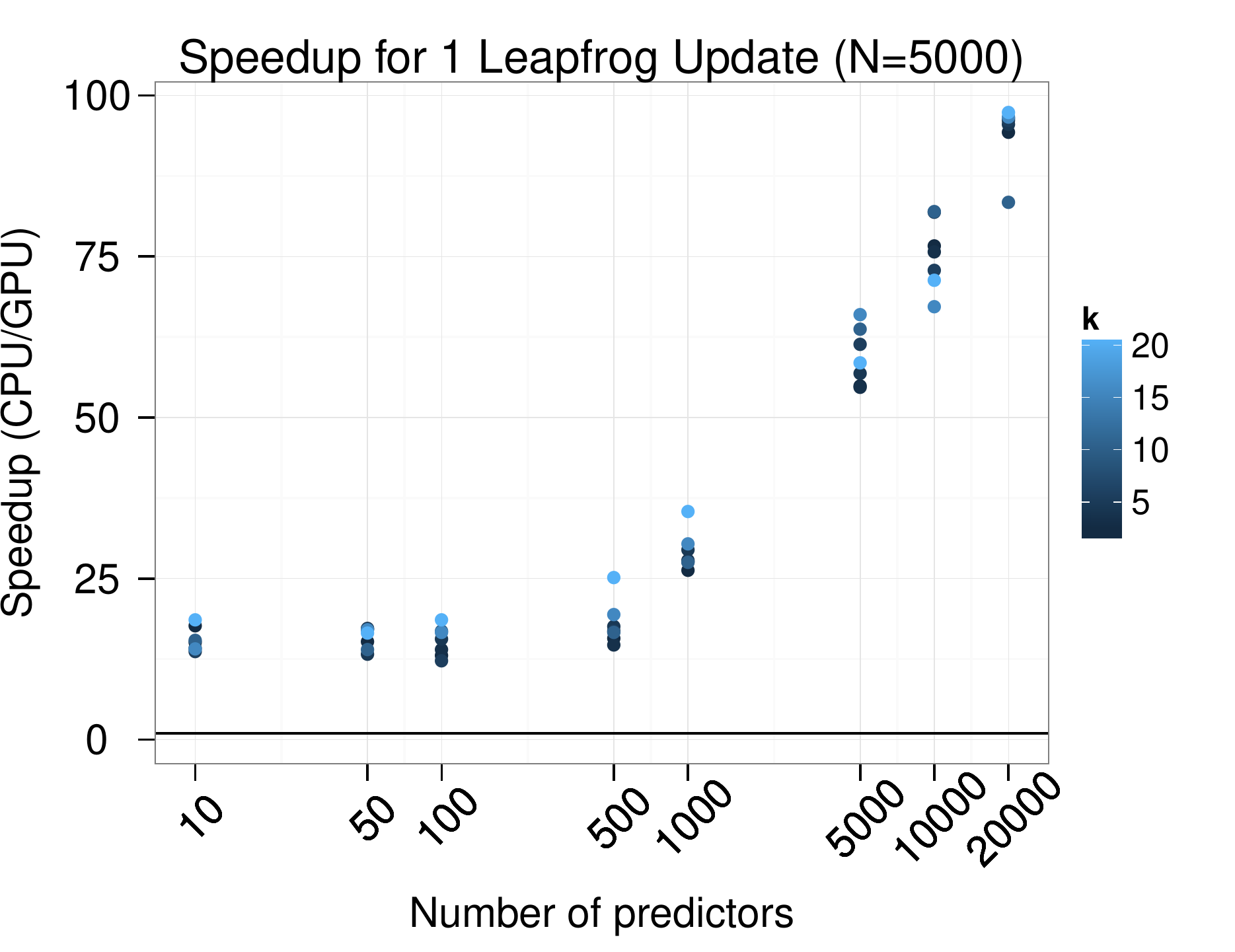}}
	\subfloat{\includegraphics[width=3.5in]{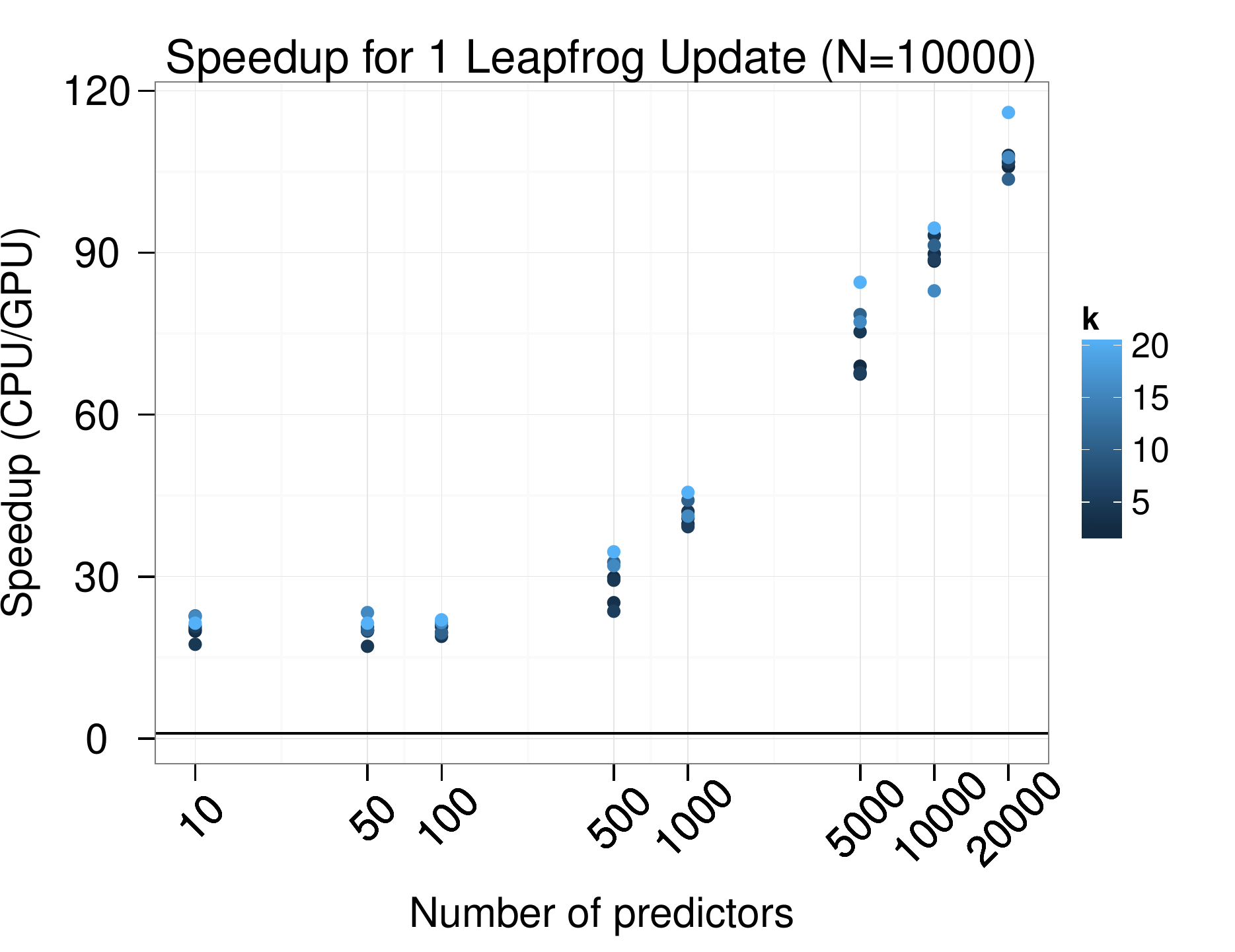}} \quad
		\caption{Timing results for 1 leap-frog update when identifiability is not enforced.} \label{fig4}
\end{figure}

However, PyCuda does not currently support slicing on GPU arrays. We could write custom CUDA C-code to only update the unconstrained coefficients, but this would greatly increase the complexity of the provided code, and users who are not familiar with GPU programming would likely get little from it. Instead, we create a $p \times k$ \emph{mask} matrix $M$, where $M_{jk} = 1 $ if $ k \neq K$ and $0$ otherwise. We element-wise multiply this mask to the result in (\ref{fullgrad}) to zero out all of the elements associated with $\beta_K$. Doing this maintains identifiability during each gradient update, but comes at cost of $O(pk)$ each time, since we have to element-wise multiply by $M$ everytime we update the gradient. However, as the results section show, this additional cost is relatively small compared to speedup offered by the GPU framework. We present this method to maintain identifiability because routinely we are interested the interpretation of each regression coefficient. However, if the task is \emph{only} prediction, this concern can be safely ignored. For completeness, we have included the results of evaluating the expressions in (\ref{logkernel}) and (\ref{gradient}) when identifiability is not a concern. These results are summarized Figures \ref{fig3} and \ref{fig4}.

\bibliographystyle{ECA_jasa}
\bibliography{ref}

\end{document}